\documentclass[aps,pra,reprint,superscriptaddress,showpacs,floatfix, longbibliography, 10pt]{revtex4-2}

\usepackage{xcolor}
\usepackage{braket}       
\usepackage{graphicx}      
\usepackage{longtable}     
\usepackage{xurl}          
\usepackage{bm}            
\usepackage{soul}
\usepackage{amsmath}
\usepackage{amssymb}
\usepackage{siunitx}
\usepackage{hyperref}
\usepackage{cleveref}
\hypersetup{
    colorlinks=true,    
    linkcolor=blue,     
    citecolor=blue,     
    filecolor=blue,     
    urlcolor=blue       
}

\begin{document}
\title{Matter wave bistability with a momentum chirped Bose-Einstein condensate}

\author{Mikhail Mamaev}
\email{mikhail.mamaev@utoronto.ca}
\altaffiliation{The first two authors contributed equally to this work.}
\affiliation{Department of Physics, University of Toronto, 60 St. George Street, Toronto ON, M5S 1A7, Canada}

\author{Nicholas Mantella}
\email          {nicholas.mantella@mail.utoronto.ca}
\affiliation{Department of Physics, University of Toronto, 60 St. George Street, Toronto ON, M5S 1A7, Canada}

\author{Noah P. Baker}
\affiliation{Department of Physics, University of Toronto, 60 St. George Street, Toronto ON, M5S 1A7, Canada}

\author{Joseph McGowan IV} 
\affiliation{Department of Physics, University of Toronto, 60 St. George Street, Toronto ON, M5S 1A7, Canada}

\author{Aephraim M. Steinberg}
\email{steinberg@physics.utoronto.ca}
\affiliation{Department of Physics, University of Toronto, 60 St. George Street, Toronto ON, M5S 1A7, Canada}

\begin{abstract}

While a Fabry-Perot (FP) cavity interferometer is a standard tool in modern optics, analogous physics can also be observed with matter waves such as Bose-Einstein condensates (BECs). Interatomic collisions in BECs can induce nonlinear features such as multi-valued transmission spectra for interferometry experiments. Observation of interaction effects in the matter-wave analogue of FP interferometry has not yet been experimentally achieved, mostly due to broadening of transmission peaks by interaction-induced shifts of cavity resonances. In this work, we theoretically show that this limitation can be overcome by engineering a momentum chirp in a BEC wavepacket. By balancing this chirp against the dynamical mean-field energy shift of the cavity during transmission, we predict enhanced resonant spectra and bistability features due to interaction-assisted tunneling that would be otherwise inaccessible. We show that this physics is readily observable with quasi-one-dimensional BECs of cold atoms colliding with a pair of light-induced Gaussian potential barriers forming the cavity. We also benchmark the requisite optimum chirp needed and characterize the contrast of near-discontinuous FP transmission spectra at an interacting resonance. Our findings enable the realization of technologies like matter-wave switching protocols, and high-precision velocity measurements; estimates of achievable sensitivity for the latter are provided.

\end{abstract}

\maketitle

\section{Introduction}

\begin{figure*}
\includegraphics[width=0.7\textwidth]{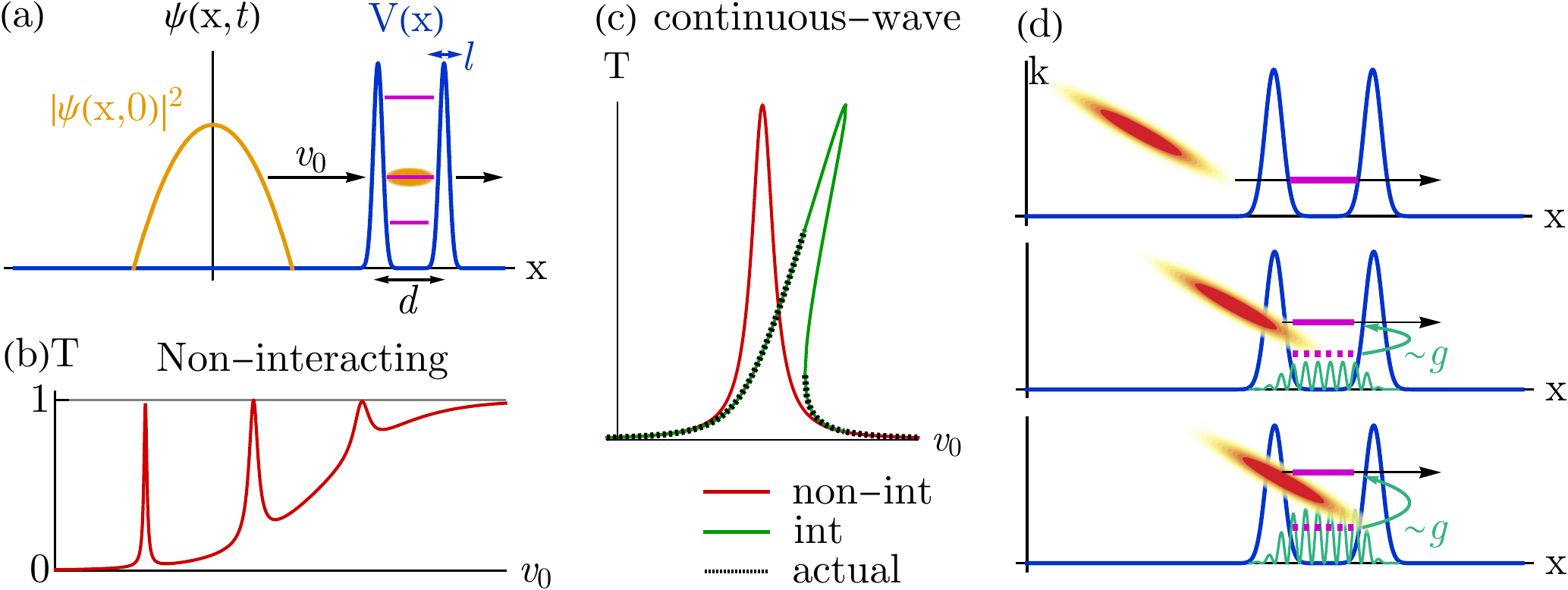}
    \caption{(a) Schematic of the system. A 1D BEC is launched at velocity $v_0$ towards a double-barrier potential forming a Fabry-Perot cavity of width $d$. Each barrier is a Gaussian of width $l$ and height $V_0$. (b) The transmission spectrum for a non-interacting BEC as a function of initial velocity. Transmission approaches $T = 1$ once the incident energy is much higher than $V_0$, but can also reach $T = 1$ near cavity resonances. (c) Cartoon schematic of how the transmission spectrum is distorted by interactions in the continuous-wave limit (an infinitely spatially extended wavepacket). The red curve is non-interacting transmission, the green curve shows the multi-valued spectrum for an interacting gas, and the black dashed curve shows the actual transmission for a dynamical experiment (which picks out only the lowest branch). The latter transmission exhibits discontinuity as a function of $v_0$. For a dynamical wavepacket of finite size, interactions can wash out these transmission features. (d) Schematic of the resonance mechanism we study: for an anti-chirped wavepacket, atoms arriving earlier are slower. If they are resonant, they fill the cavity and shift the resonance via their mean-field energy. For the right amount of anti-chirp, this shift enables later, faster atoms to also be resonant.
}
\label{fig: schematic}
\end{figure*}

The use of ultracold matter available in experimental atomic physics has been an extraordinary tool for the investigation and simulation of fundamental quantum phenomena. Bose-Einstein condensates (BECs) in particular have enabled studies of matter-wave optics. Through the use of optical and magnetic potentials, matter waves formed by a BEC can be manipulated in analogous ways to light fields. This control has enabled fruitful studies of long-standing questions in both practical sensing applications and foundational quantum mechanics. One such area of study is the dynamics of quasi-one-dimensional (1D) BECs moving through a potential landscape. In particular, the tunneling of a BEC through a single Gaussian barrier has been used to gain insight into the long-standing question of the duration of a tunneling event~\cite{ramos2020measurement}. There have also been experimental studies of Bragg diffraction in matter waves realized by BECs propagating in waveguides~\cite{fabre2011realization}. More broadly, the topic is of relevance to atomtronics and transistor design~\cite{amico2022colloquium, anderson2021matter, dowarah2025coherent}, as well as theoretical ideas such as interaction-induced asymmetry of tunneling under spatial reflection~\cite{lindberg2023asymmetric}. With the aid of tools like spatial light modulators, it is possible to create almost arbitrary potential barriers for studying fundamental physics like parity-time (PT) symmetry breaking~\cite{soley2023experimentally}. A particularly interesting potential to consider is a pair of Gaussian barriers, closely spaced, which form a coherent matter-wave analogue of a Fabry-Perot (FP) cavity. Specifically, in this work we consider the setup depicted in Fig.~\ref{fig: schematic}(a), where a spatially-finite BEC is pushed through a FP.

The possibility of realizing a matter-wave FP interferometer has been discussed theoretically for several decades~\cite{carusotto2001nonlinear, paul2005nonlinear, damon2015band, manju2020atomic}. Matter waves replace the usual incoming light waves, and laser-induced potential barriers act as cavity mirrors. Like conventional cavities, the transmission spectrum through the FP for an ideal non-interacting incoming wavepacket exhibits Breit-Wigner resonances at energies corresponding to quasi-bound states of the FP~\cite{ricco1984physics, ferry1999transport}, as depicted in Fig.~\ref{fig: schematic}(b). Resonant transmission enables applications such as momentum-filtering (since narrow resonances yield output matter waves with very low momentum uncertainty)~\cite{smith2019engineering}, a potential source for an atom laser, and even gravimetry measurements of acceleration in low-gravity environments~\cite{schach2022tunneling, perumbil2025theoretical}; the latter has seen recent experimental progress~\cite{gaaloul2022space}. FP transmission resonances have been observed in the case of a non-interacting Bose-Einstein condensate of $^{39}$K~\cite{eid2024ultracold}. There is also a recent experiment that uses non-interacting atoms confined inside a matter-wave FP to study the effects of curved spacetime~\cite{tanlimco2026matter}. However, experimental realizations displaying features exclusive to an interacting BEC remain lacking. 

It is well known that in the mean-field description, an interacting BEC can exhibit many of the nonlinear effects present in $\chi^{(3)}$ nonlinear optics~\cite{paul2005nonlinear, paul2007nonlinear}. In particular, one can find a discontinuity in the transmission spectrum as a function of the incident wavepacket velocity, as depicted qualitatively in Fig.~\ref{fig: schematic}(c). While non-interacting particles feature a conventional Lorentzian resonance lineshape, the interacting case exhibits a distortion in the spectrum leading to multi-valued transmission. A dynamical tunneling process tends to follow the lowest transmission branch, leading to discontinuity when degeneracy first appears as a function of incident velocity $v_0$ (for repulsive interactions). Such discontinuous transmission features can enable an effective matter-wave ``switch" or low-pass filter. While these features were theoretically predicted for continuous-wave beams of particles relevant to e.g. quantum dot systems, analogous features in cold atom platforms with spatially-finite BECs have not yet been observed.

There are two main difficulties in observing matter-wave FP resonances. The first is the narrow initial momentum width required. The energy scale of an FP cavity potential formed by far-detuned Gaussian laser beams is constrained by the diffraction limit, which makes it difficult to create a cavity with a length smaller than a few microns for optical-frequency beams. Resolving the resonances with BECs of typical atoms like $^{87}$Rb requires the initial wave-packet to have kinetic temperature on the order of hundreds of pK before studying collision physics. This level of cooling is difficult, although it has entered the realm of experimental feasibility in several labs~\cite{kovachy2015matter, deppner2021collective, gaaloul2022space}.

The second difficulty, unique to the interacting case, is the shift of the cavity resonances due to the interaction energy of the atoms while they are inside. For a single atom, sufficiently strong interactions can lead to blockade effects~\cite{carusotto2001nonlinear} where existing population can block transmission for subsequent atoms. Likewise, for a \textit{transient} scattering process with a many-atom BEC, the mean-field shift from earlier-arriving atoms can cause later atoms to become off resonant as the process occurs even if the whole wavepacket starts on-resonance (we distinguish different atoms in the wavepacket as a short hand, but generally refre to the spatial probability amplitude of the condensate). Prior work found that FP resonances for transient wavepackets were washed out even for interactions with low collisional scattering lengths~\cite{manju2020atomic}, with no signatures of $\chi^{(3)}$-nonlinear transmission spectra. It is this problem that the present work aims to address.

We show that it is possible to reveal resonances in FP cavities via interaction-assisted tunneling. Our approach is to engineer a momentum anti-chirp in the initial wavepacket using a matter-wave lensing technique before colliding it with the cavity. This anti-chirp generates a negative correlation between the wavepacket's starting position and velocity in phase space. We find that the right ``magic" amount of chirp can exactly cancel the increasing interaction shift of the cavity resonance as atoms enter it, allowing more atoms to remain resonant throughout the scattering event, as depicted in Fig.~\ref{fig: schematic}(d). While normally any chirp or anti-chirp increases momentum uncertainty of a wavepacket and washes out FP resonances, a ``magic" chirp instead amplifies them to the point where they are observable under realistic experimental conditions. We find the optimal chirp under experimentally reasonable parameters for a state-of-the-art platform employing $^{87}$Rb atoms. We also show how the visibility of the resonances can be improved if it is possible to tune the interaction strength of the BEC, using a Feshbach resonance (FR). We further characterize the nearly-discontinuous transmission spectrum that can be observed with magically chirped wavepackets. The near-discontinuity enables high-precision measurement of incident velocity, for which we provide an estimate of the achievable sensitivity.

The remainder of this paper is structured as follows. In Section~\ref{Sec 2} we discuss the simulation methods used, including state preparation, matter-wave lensing, and collisions with FP potentials. In Section~\ref{Sec 3} we find the optimal momentum chirp for a variety of interaction strengths, investigate the dependence of resonance sharpness on interaction strength and chirp angle, and estimate the achievable sensitivity of velocity measurements. In Section~\ref{Sec 4}, we present some approximate analytic results to complement our numerical simulations. Lastly, in Section~\ref{Sec 5} we conclude and summarize our results.

%%%%%%%%%%%%%%%%%%%%%%%%%%%%%%
\section{Methods}\label{Sec 2}
%%%%%%%%%%%%%%%%%%%%%%%%%%%%%%

We theoretically consider a spinless BEC confined to one dimension $x$, as depicted in Fig.~\ref{fig: schematic}(a), with strong transverse confinement along the directions $y$ and $z$ (i.e. a cigar-shaped trap). There is an additional external potential $V(x)$ along the $x$ direction used to generate barriers and manipulate the condensate. The dynamics of the condensate wavefunction $\psi(x,t)$ are modeled at the mean-field level by the standard Gross-Pitaevskii equation (GPE) reduced to 1D~\cite{salasnich2002effective},
\begin{equation}
\begin{aligned}
\label{eq_GPE}
i \hbar\frac{\partial}{\partial t} \psi(x,t) = -\frac{\hbar^2}{2m} \frac{\partial^2}{\partial x^2}\psi(x,t) + V(x)\psi(x,t)&\\
+ g |\psi(x,t)|^2 \psi(x,t)&.
\end{aligned}
\end{equation}
Here $m$ is the atomic mass and $g$ is the effective interaction strength in 1D, $g>0$ (repulsive) unless indicated otherwise. This effective interaction strength is obtained by integrating out the strong harmonic confinement along the transverse dimensions with trapping frequency $\omega_{\perp}$, yielding $g = 2N \hbar a_s \omega_{\perp}$,
with $N$ the number of atoms and $a_s$ the collisional scattering length. This approximation for $g$ holds as long as the transverse confinement harmonic oscillator energy $\hbar \omega_{\perp}$ is large compared to the interaction energy; more sophisticated approximations exist~\cite{mateo2008effective,manju2020atomic}, though we do not expect such details to qualitatively affect our results. Note the interaction strength $g$ is proportional to atom number $N$ because we choose the condensate wavefunction to be normalized to $\int dx |\psi(x)|^2 = 1$. In what follows, we will assume the following parameters unless otherwise specified: $N=3000$ atoms, $m=87$ amu, $a_s=96 a_0$ (where $a_0$ is the Bohr radius) and $\omega_{\perp}/(2\pi)=250$ Hz, hence $g/(2\pi\hbar) = 7.6$ Hz$\times $mm.

\begin{figure*}
\centering
    \includegraphics[width=1\textwidth]{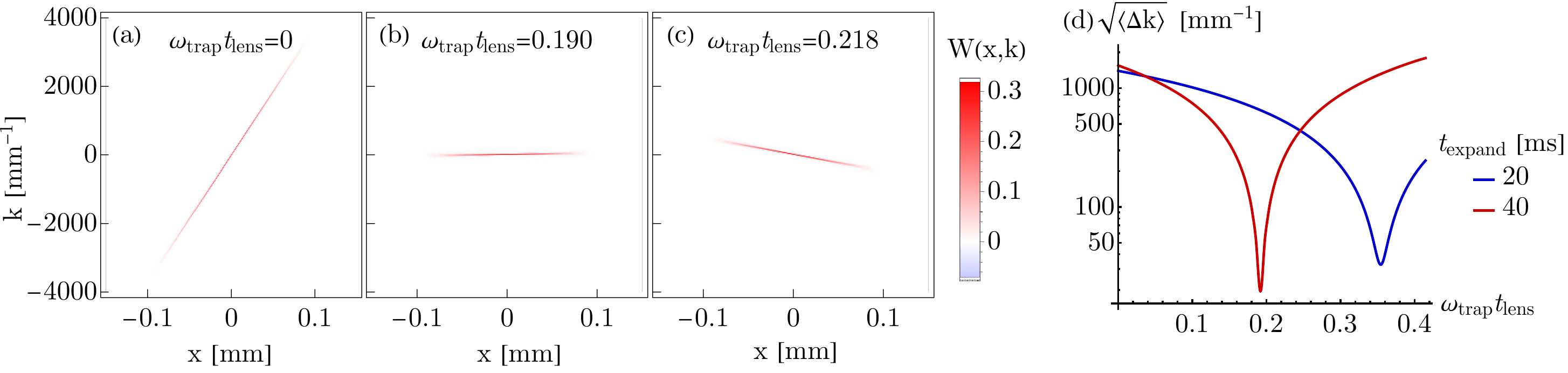}
    \caption{(a) Wigner function of the condensate after $t_{\mathrm{expand}}=40$ ms of free expansion. (b-c) Wigner function of the condensate after applying a delta-kick with the harmonic trap, rotating the phase-space distribution by an amount $\omega_{\mathrm{trap}}t_{\mathrm{lens}}$. Panel (b) shows the kick with lensing time $t_{\mathrm{lens}} = t_{\mathrm{opt}}$ (given by $t_{\mathrm{opt}} = 1.38$ ms for these parameters) that yields a momentum-minimized state, while panel (c) shows an over-rotated (i.e. anti-chirped) state. The chirp angle for these is given by $\theta = \omega_{\mathrm{trap}}(t_{\mathrm{lens}} - t_{\mathrm{opt}})$, which is $\theta = -0.19, 0, 0.28$ for panels (a-c) respectively. (d) Momentum variance of the state versus rotation angle $\omega_{\mathrm{trap}}t_{\mathrm{lens}}$ for different amounts of initial expansion time before the kick.}
    \label{fig: dkc}
\end{figure*}

\subsection{Initial state and matter-wave lensing}

Before any barrier collisions, we first describe the initial state of the system. We assume the atomic cloud is loaded into a longitudinal harmonic trapping potential $V(x) = V_{\mathrm{trap}}(x)$ given by:
\begin{equation}
V_{\mathrm{trap}}(x) = \frac{1}{2}m \omega_{\mathrm{trap}}^2 x^2,
\end{equation}
and cooled into the ground-state, with $\omega_{\mathrm{trap}}$ the longitudinal trap frequency (set to $\omega_{\mathrm{trap}}/(2\pi) = 22$ Hz hereafter). We numerically obtain the ground-state wavefunction in the trap via imaginary-time evolution of the GPE equation, starting from a non-interacting harmonic oscillator ground state of $V_{\mathrm{trap}}(x)$. The resulting ground state is approximately equal to a Thomas-Fermi (TF) distribution with TF radius $r_{\mathrm{TF}}=14$ $\mu$m for our choice of parameters.

Our goal is to resolve narrow energetic features like cavity resonances via barrier-collisions. To this end, naively one must prepare an initial wavefunction which has minimized momentum uncertainty. For an interacting BEC confined to a harmonic trap this can be difficult to achieve, as there is significant mean-field interaction energy for $g \neq 0$ which will convert to kinetic energy and broaden the wavepacket (since one does not keep the BEC trapped during the collision process). One could turn the interactions off with a Feshbach resonance. However, we seek to derive a theory applicable to a general class of atoms including those such as $^{87}$Rb. This atom lacks an easily accessible Feshbach resonance, yet remains a prime candidate for BEC experiments due to its simplicity and favorable scattering properties for evaporative cooling. We can minimize the momentum uncertainty of the initial wavepacket for non-Feshbach atoms via the technique of delta-kick cooling (DKC)~\cite{chu1986proposal}.

In a DKC protocol, the BEC is first released to undergo free expansion. Specifically, we set $V(x) = 0$ and numerically time-evolve the condensate under just the kinetic and interaction terms of the GPE for a time $t_{\mathrm{expand}}$, causing it to spread and convert mean-field interaction energy into kinetic energy. This conversion generates significant broadening in the condensate's momentum uncertainty. Fig.~\ref{fig: dkc}(a) shows the Wigner function for the condensate wavefunction after $t_{\mathrm{expand}}=$40 ms. This Wigner function retains the elliptical shape of the initial state~\cite{castin1996bose, kagan1996evolution}, but becomes squeezed along a particular angle in phase space (and anti-squeezed along the perpendicular direction). The specific squeezed quadrature is a combination of $\hat{x}$ and $\hat{k} = \hat{p}/\hbar$ (for momentum $\hat{p}$) that depends on the expansion time and system parameters.

The next step in DKC is to minimize the momentum uncertainty via rotation of the squeezed quadrature, which can be accomplished by turning the harmonic trapping potential $V(x) = V_{\mathrm{trap}}(x)$ back on for a time $t_{\mathrm{lens}}$. Doing so rotates the phase-space distribution of the condensate by an angle $\omega_{\mathrm{trap}}t_{\mathrm{lens}}$. Fig.~\ref{fig: dkc}(a-c) shows this distribution for different lensing times. There is a particular angle at which the momentum uncertainty is minimized. As shown in Fig.~\ref{fig: dkc}(d), the minimum achievable uncertainty grows smaller for longer expansion times as more mean-field energy is expended to squeeze the state. It is also possible to induce a momentum chirp (anti-chirp) by under (over) rotating the momentum distribution. We define the ``chirp angle" as $\theta=\omega_{trap}(t_{lens}-t_{\mathrm{opt}})$, where $t_{lens}$ is the variable lensing time, and $t_{\mathrm{opt}}$ is the lensing time required to minimize the momentum width of the wave-packet.

Note that while an ideal harmonic potential has no limit to how narrow the momentum width can be, in practice, there are limitations due to the anharmonic nature of the Gaussian potentials generated by far-detuned lasers. Such experimental imperfections are investigated in Appendix~\ref{app_Imperfections}. We also note that there are several other approaches to reducing initial momentum uncertainty without a Feshbach resonance, which include the use of motion in linear potentials~\cite{damon2014reduction} and the use of multiple kick pulses rather than just one~\cite{neeraj2026improved}. Note also that for a finite time $t_{\mathrm{expand}}$ before exploring collision physics, there is still some amount of residual mean-field interaction energy, which will cause the wavepacket to continue to expand in $k$-space as it flies towards the barrier. However, this additional expansion is negligible in our subsequent results.

%%%%%%%%%%%%%%%%%%%%%%%%
\subsection{Fabry-Perot barrier collision}
%%%%%%%%%%%%%%%%%%%%%%%%

We can now numerically study scattering of the lensed condensate through a FP potential $V(x) = V_{\mathrm{FP}}(x)$. We model this potential as two Gaussian repulsive barriers,
\begin{equation}
V_{\mathrm{FP}}(x) = V_0 e^{-2(x-x_0-d/2)^2/l^2} + V_0 e^{-2(x-x_0+d/2)^2/l^2}.
\end{equation}
Here $V_0$ is the barrier height, $x_0$ is the center of the FP (assumed to be outside the condensate's initial wavefunction), $d$ is the FP cavity length, and $l$ is the width of each barrier ($1/e^2$ radius). In all results hereafter we take $x_0 = 150$ $\mu$m, $d = 5$ $\mu$m, $l = 1.3$ $\mu$m; these numbers closely match achievable values in a state-of-the-art platform using $^{87}$Rb atoms~\cite{spierings2024spin}.

The lensed cloud is given a momentum kick $k_{0}$, corresponding to a velocity $v = v_0 \equiv \frac{\hbar}{m}k_0$, pushing it  towards the potential:
\begin{equation}
\psi(x) \to e^{i k_0 x}\psi(x).
\end{equation}
We numerically time-evolve this pushed state under the GPE with $V_{\mathrm{FP}}(x)$ for a sufficiently long time that the collision finishes happening (typically requiring $t v_0 \geq 2 x_0$). The transmission probability is obtained from the modulus squared of the wavefunction in the region beyond the cavity,
\begin{equation}
T = \int_{x_0+d/2}^{\infty}dx |\psi(x,t)|^2\bigg|_{t\gtrsim 2x_0/v_0}.
\end{equation}
The transmission spectrum for the ideal, non-interacting case is qualitatively depicted in Fig.~\ref{fig: schematic}(b). Transmission approaches 100$\%$ when incident energy matches one of the cavity resonances. For high enough incoming velocity, transmission also approaches 100$\%$ independent of resonance since the kinetic energy becomes much larger than the barrier height and classical transmission becomes possible.

In Fig.~\ref{fig: schematic}(c) we also qualitatively sketch the difference in transmission spectra for interacting and non-interacting particles for a continuous-wave setup, where the spatial extent of the wavepacket is so large that the system dynamics are essentially time-independent. In this case, it is known that interactions cause the spectrum to become distorted and multi-valued~\cite{paul2005nonlinear, paul2007nonlinear}. A dynamical experiment measuring tunneling will only observe the lowest branch of this multi-valued spectrum, leading to discontinuity in the transmission as a function of $v_0$. For an attractive interaction $g < 0$, the discontinuity would be equivalent but skewing left instead of right (i.e towards lower $v_0$). In the case of a finite wave-packet, such features are normally broadened when the momentum width of the incoming BEC is larger than the width of the resonance. This leads to a reduction in the maximum transmission, and if the momentum width is broad enough the resonances are washed out completely rather than exhibiting any non-trivial or discontinuous features. The goal of our work is to find a regime where such features can still be observed.

%%%%%%%%%%%%%%%%%%%%%%%%%%%
\section{Transmission probability results}\label{Sec 3}
%%%%%%%%%%%%%%%%%%%%%%%%%%%
%%%%%%%%%%%%%
\subsection{Optimal momentum chirps}
%%%%%%%%%%%%%

\begin{figure*}
\centering
    \includegraphics[width=1\textwidth]{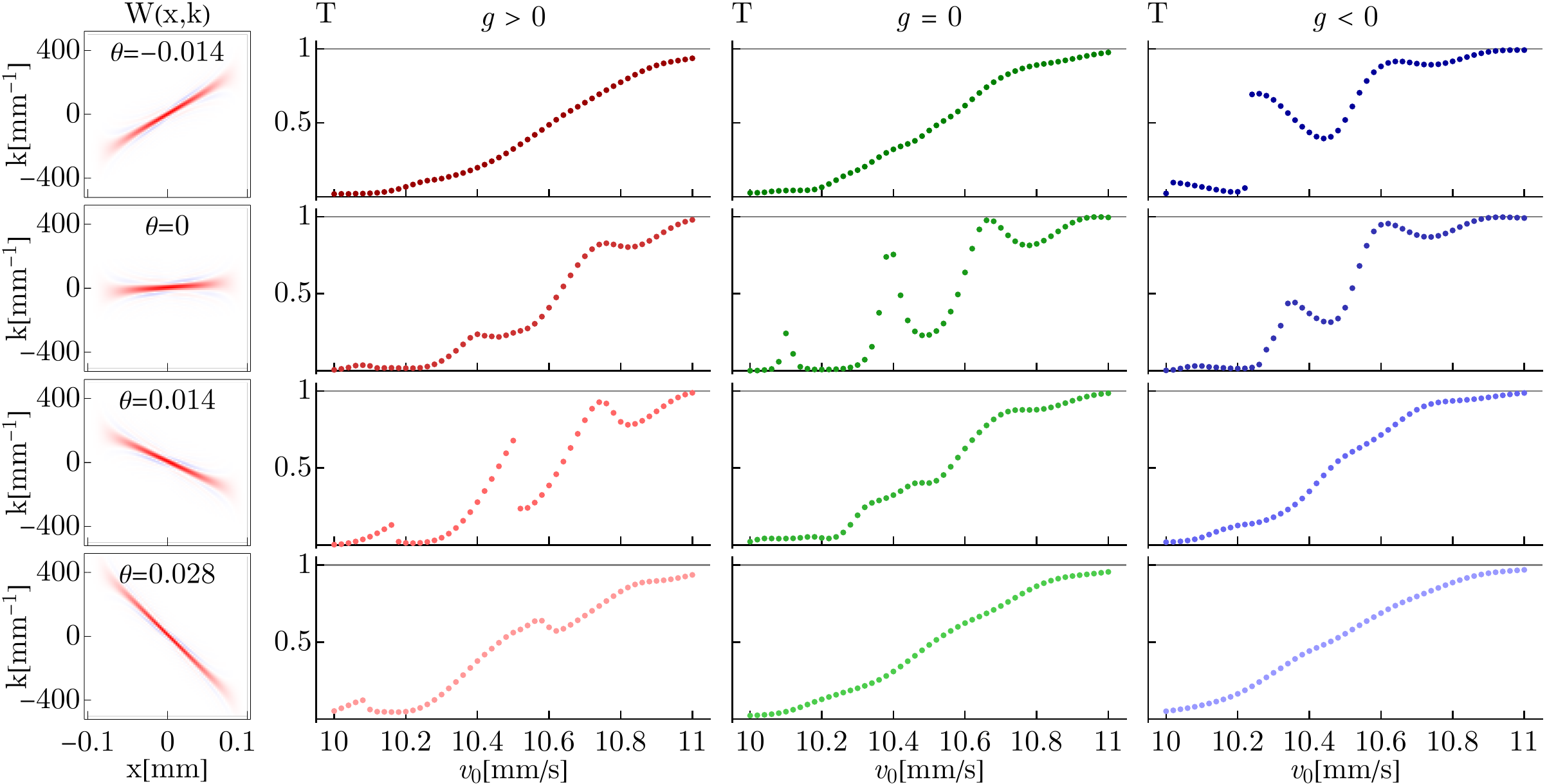}
    \caption{Transmission probability $T$ versus incident velocity $v_0$ of the wavepacket for repulsive $g >0$, non-interacting $g = 0$ and attractive $g < 0$ interactions. We also vary the initial chirp angle $\theta$; the initial state's Wigner function for each row of panels is shown in the corresponding left panel (before being kicked towards the cavity, hence centered at momentum $k = 0$ rather than $k = k_0$). For all simulations, the initial state is prepared assuming a repulsive interaction $g > 0$ and $t_{\mathrm{expand}} =40$ ms of expansion time; the interaction strength $g$ is quenched to zero or negative values for the corresponding columns of panels during the collision dynamics only.}
    \label{fig: chirpvsint}
\end{figure*}

In Fig.~\ref{fig: chirpvsint}, we show the transmission probability $T$ versus the incident velocity $v_0$ for non-interacting ($g = 0$), repulsive ($g > 0$) and attractive ($g < 0$) gases. For simplicity, all simulations use the same initial condition obtained via the methods described in the previous section for a repulsive $g >0$, and the value of $g$ is switched for the collision process only. In the case of a non-interacting BEC, optimal resonance visibility occurs for minimum initial momentum uncertainty hence chirp angle $\theta = 0$. Any chirp ($\theta < 0$) or anti-chirp ($\theta > 0$) means increased initial momentum width of the incoming wave-packet and broadening of the resonances as expected.

For an interacting BEC, we find that starting with minimum initial momentum uncertainty $\theta = 0$ yields broadened resonant peaks, especially for repulsive interactions $g > 0$. This occurs due to the aforementioned dynamical mean-field shift of the cavity resonance. If the earliest atoms arriving at the cavity are resonant with a quasi-bound state, they fill the cavity and shift the bound state's energy up or down (depending on the sign of $g$). This shift prevents subsequent atoms from entering the cavity, since they have the same energy as earlier ones for a momentum-minimized initial state with no chirp, and are no longer resonant.

As seen in Fig.~\ref{fig: chirpvsint}, for the right ``magic" amount of anti-chirp $\theta >0$ for repulsive interactions, or the right amount of chirp $\theta < 0$ for attractive interactions, the visibility of the resonances improves dramatically and the discontinuous nature of the transmission spectrum can be seen. Despite the naively increased momentum width of chirped or anti-chirped wavepackets, their spatiotemporal correlations mediate better resonant transmission. We attribute this improvement to the aforementioned effect depicted in Fig.~\ref{fig: schematic}(d). Earlier atoms shift a bound state's energy by an amount that matches the energy difference with later atoms caused by the (anti-) chirp, allowing a larger fraction of the cloud to remain on-resonant for the duration of the interaction. If the chirp is too strong, this cancellation fails and the resonances return to being washed out. We thus find that there is a ``magic" (anti-)chirp angle that maximizes resonance visibility. In Section~\ref{Sec 4} we provide a simplified analytic approach that verifies the origin of this effect.

We also note that the position of the resonance shifts to higher/lower velocity with different chirp angles for repulsive/attractive interactions. This can also be understood from the interaction-mediated picture. For e.g. repulsive interactions, the mechanism we describe requires the earliest arriving atoms to be resonant. With a stronger chirp, the cloud has a broader momentum distribution, so we need a larger difference in velocity to ensure its ``edge" in phase-space coincides with the resonance energy.

%%%%%%%%%%%%%%
\subsection{Interaction-dependent contrast}
%%%%%%%%%%%%%%

As seen in Fig.~\ref{fig: chirpvsint}, resonances in the interacting case can have a large change in transmission over a relatively small change in incident velocity. This effect is of interest for future experiments involving interacting matter-wave FPs for creating matter-wave switches, or for highly sensitive velocity measurements. The latter is especially relevant to certain types of gravimetry measurements, such as measurement of acceleration in microgravity~\cite{perumbil2025theoretical}. It is therefore of interest to find the interaction strength and chirp angle that yield the maximum possible change in transmission for a given change in incident velocity.

We first perform a coarse-grained analysis of achievable sensitivity, which assumes the initial velocity $v_0$ can only be defined up to a finite resolution $\Delta v_0$. We define the contrast,
\begin{equation}
    C(v_0)=\left|\frac{T(v_0) - T(v_0 + \Delta v_0)}{T(v_0) + T(v_0 + \Delta v_0)}\right|,
\end{equation}
where $T(v_0)$ is the transmission for initial velocity $v_0$, and $\Delta v_0$ is a fixed velocity spacing (in our case we choose $\Delta v_0 = 5 \mu$m/s). This contrast characterizes how big a jump in transmission one can resolve given the fixed $\Delta v_0$. Note that for a spatially-finite wavepacket, the transmission curve is continuous (true discontinuity only occurs in the continuous-wave limit), meaning that the contrast will approach zero for $\Delta v_0 \to 0$. However, for practical setups, a coarse-grained analysis using a finite $\Delta v_0$ is still relevant due to experimental uncertainty of the imparted momentum kick.

In Fig.~\ref{fig: optcont}(a), we plot the maximum contrast achievable for our choice of velocity resolution versus interaction strength (normalized by our conventional choice of $g=g_0$ with $g_0/(2\pi \hbar) = 7.6$ Hz$\times$mm using scattering length of $96 a_0$ for $^{87}$Rb). This maximum contrast is obtained by maximizing over all initial velocities $v_0$ and all chirp angles, and generally occurs at the near-discontinuous portion of the transmission spectra. The optimal chirp angle $\theta$ that yields maximum contrast for these results is shown in Fig.~\ref{fig: optcont}(b).

An important feature of these results is that maximum contrast for an attractive interaction strength is always larger than for the corresponding repulsive interaction strength of the same magnitude. This is because non-resonant transmission increases at higher energies, meaning that shifting a resonance to lower energy (which attractive interactions do) leads to a larger change in transmission over the same change in incident velocity. This is significant motivation for using an atomic species with an experimentally practical Feshbach resonance such as $^{39}$K. As a caveat, pushing to interaction strengths too strong will lead to breakdown of the mean-field Gross-Pitaevskii description.

%%%%%%
\subsection{Velocity measurement uncertainty}
%%%%%%

While the above analysis studied a coarse-grained transmission spectrum, we can also consider the limit where we seek to measure arbitrarily small changes in velocity $\Delta v_0$. The local uncertainty in measuring $v_0$ from measurements of $T(v_0)$ is given by,
\begin{equation}
\sigma_{v_0}=\frac{\sigma_{T}(v_0)}{\sqrt{M} \left|\frac{\partial T(v_0)}{\partial v_0}\right|},    
\end{equation}
where $M$ is the number of runs (i.e. number of times the experiment is repeated), and $\sigma_T(v_0)$ is the uncertainty in measurements of transmission $T$ for the full cloud; we assume for simplicity that these are velocity-independent $\sigma_T(v_0) = \sigma_T$, meaning the only relevant parameter is the derivative of transmission with respect to initial velocity $\left|\frac{\partial T(v_0)}{\partial v_0}\right|$.

\begin{figure*}
\centering
    \includegraphics[width=0.95\textwidth]{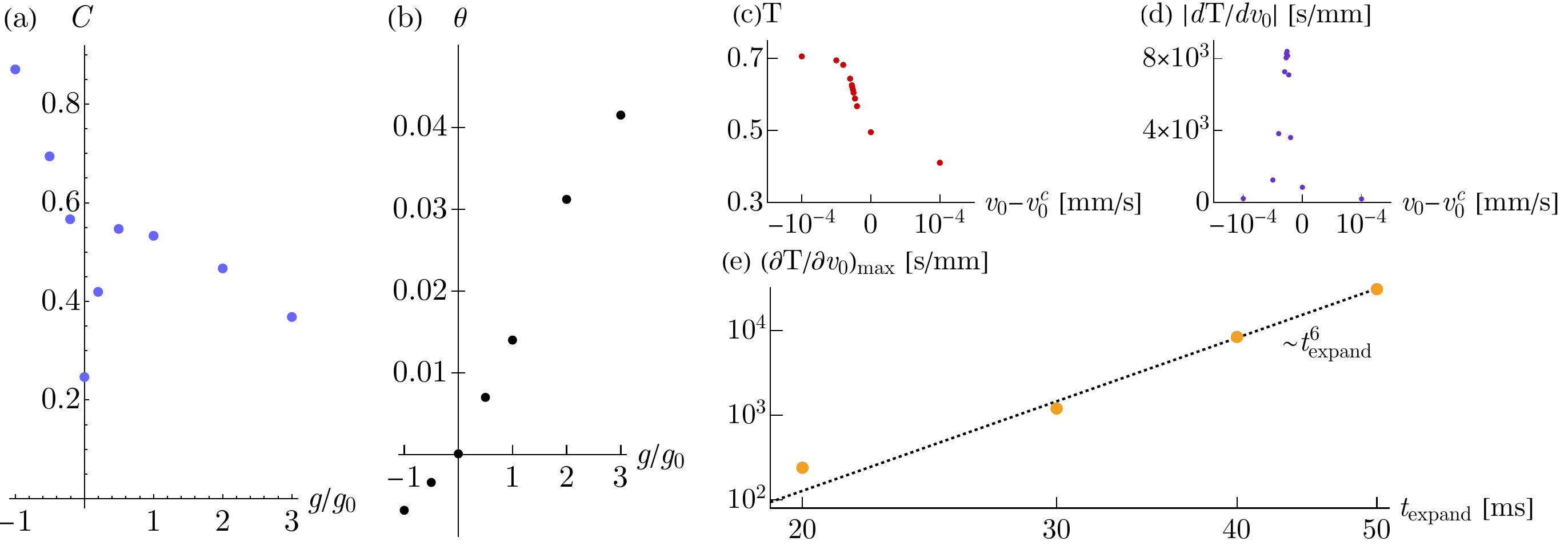}
    \caption{(a) Maximum contrast of the transmission spectra from Fig.~\ref{fig: chirpvsint} with a coarse-grained velocity resolution of $\Delta v_0 = 5 \mu$m/s, as a function of interaction strength $g$. The interaction strength is normalized by the usual value $g_0/h = 7.6 $ Hz $\times$mm used in all prior simulations (corresponding to a positive scattering length of $96 a_0$ for $^{87}$Rb). Each point is shown for the chirp angle $\theta$ that yields maximum contrast. (b) Optimal chirp angle for maximum contrast. (c) Transmission spectrum for $g = g_0$ near the near-discontinuous portion of the most visible resonance at velocity $v_0^c = 10.5106$ mm/s. (d) Derivative of the transmission with respect to velocity $v_0$. (e) Maximum derivative as a function of free expansion time $t_{\mathrm{expand}}$ for the cloud before doing FP interferometry.}
    \label{fig: optcont}
\end{figure*}

In Fig.~\ref{fig: optcont}(c), we plot the transmission spectrum zoomed in to a near-discontinuity. Fig.~\ref{fig: optcont}(d) shows the corresponding derivative $\left|\frac{\partial T(v_0)}{\partial v_0}\right|$ in this region. The maximum derivative will depend strongly on the spatial extent of the cloud; one that has been allowed to expand more (longer $t_{\mathrm{expand}}$) will have a more narrow profile in phase-space, and yield better sensitivity assuming the optimal chirp angle for it is used. Fig.~\ref{fig: optcont}(e) shows the maximum derivative as a function of expansion time. We find a scaling of $\sim t_{\mathrm{expand}}^6$, which is equivalent to a scaling of $\sim E_{\mathrm{int}}^{-6}$, with $E_{\mathrm{int}} = \frac{g}{2}\int dx |\psi(x,0)|^4$ the residual mean-field interaction energy of the condensate after expansion and before the collision. This scaling suggests that considerable velocity sensitivity advantage can be obtained from pushing interacting matter-wave FP BEC experiments to larger trap and cloud sizes. For a setup with e.g. $\sigma_T = 0.1$ (10\% transmission resolution) and $|dT/dv_0| = 10^{4}$s/mm (what we find for $t_{\mathrm{expand}}=40$ ms), one could theoretically expect $\sigma_{v_0}= 10^{-5}$ mm/s, or sub-$\mu$m/s resolution.

To be more concrete, we envision a measurement strategy where one repeats an experiment for increasing velocities $v_0$ (shifting them up from an unknown initial value by kicking the condensate) until a sharp drop in transmission is found. One can search efficiently by taking an interval between two velocities far from resonance and continually sub-dividing the interval to isolate the sharp drop, which should allow one to reach the region with steep derivative exponentially quickly in number of experimental runs. At this point, further measurements can be made with a sensitivity of $\sigma_{v_0}$. This approach mitigates issues like small dynamic range; if $T$ was uniform outside the area with sharp derivative, or exhibited phase-winding issues, it would be more challenging to make use of the high-derivative sensitive region.

One can also consider the effects of uncertainty in the initial atom number $N$, for which the velocity sensitivity can be written as,
\begin{equation}
\sigma_{v_0}=\frac{\sqrt{\sigma_{T}^2+|\frac{\partial T}{\partial N}|^2\sigma_{N}^2}}{\sqrt{M} \left|\frac{\partial T(v_0)}{\partial v_0}\right|},
\end{equation}
with $\sigma_N$ the uncertainty in $N$ before the collision occurs. While transmission probability is always between 0 and 1 by normalizing to atom number, different $N$ yield different transmission profiles, since the resonance position shifts with $g$ which in turn depends on $N$. Too big a discrepancy in $N$ from run to run in the experiment can lead to some runs missing the high-derivative sensitive region. Numerical tests for the same expansion time $t_{\mathrm{expand}} = 40$ ms near the highest-derivative velocities yield $\left|\frac{\partial T}{\partial N}\right|\sim 10^{-3}$, which means that for $\sigma_N \sim 10^2$ (fluctuations on the order of hundreds of atoms), the noise contribution $|\frac{\partial T}{\partial N}|^2\sigma_{N}^2$ is of the same order as $\sigma_{T}^2$ and we can anticipate the same promising sub-$\mu$m precision.

%%%%%%%%%%%%%%%%%%%%%%%%%%%%%%%%%%
\section{Analytics}\label{Sec 4}
%%%%%%%%%%%%%%%%%%%%%%%%%%%%%%%%%%

The preceding sections have demonstrated that the transmission spectrum for a FP matter-wave cavity can be significantly sharpened with a ``magic" choice of anti-chirp (for $g>0$) or chirp (for $g < 0$) which cancels the time-dependent energy shift of the cavity resonance during the collision. While the GPE simulations verify this effect as genuine, we also provide a simplified analytic model that captures the effect more clearly.

The details of our derivation are in Appendix~\ref{app_Analytics}. The main methodology extends the approach in Ref.~\cite{paul2007nonlinear}, and can be summarized as follows. We assume that only one cavity bound state is relevant to the transmission properties. The interactions are assumed to only be non-negligible for the bound state, omitting any collisional effects outside the cavity. The wavefunction is expanded in terms of incoming scattering states left of the cavity, the one bound state, and outgoing scattering states right of the cavity. The cavity barriers are treated as spatial boundaries with an effective energy-dependent coupling matrix element $V_E$ in terms of scattering state energy $E$ between scattering and bound states obtained via a Bardeen surface integral. Under these assumptions, one can derive an equation of motion for the wavefunction probability amplitude of being in the bound state $B(t)$:

\begin{figure}
\centering
\includegraphics[width=0.5\textwidth]{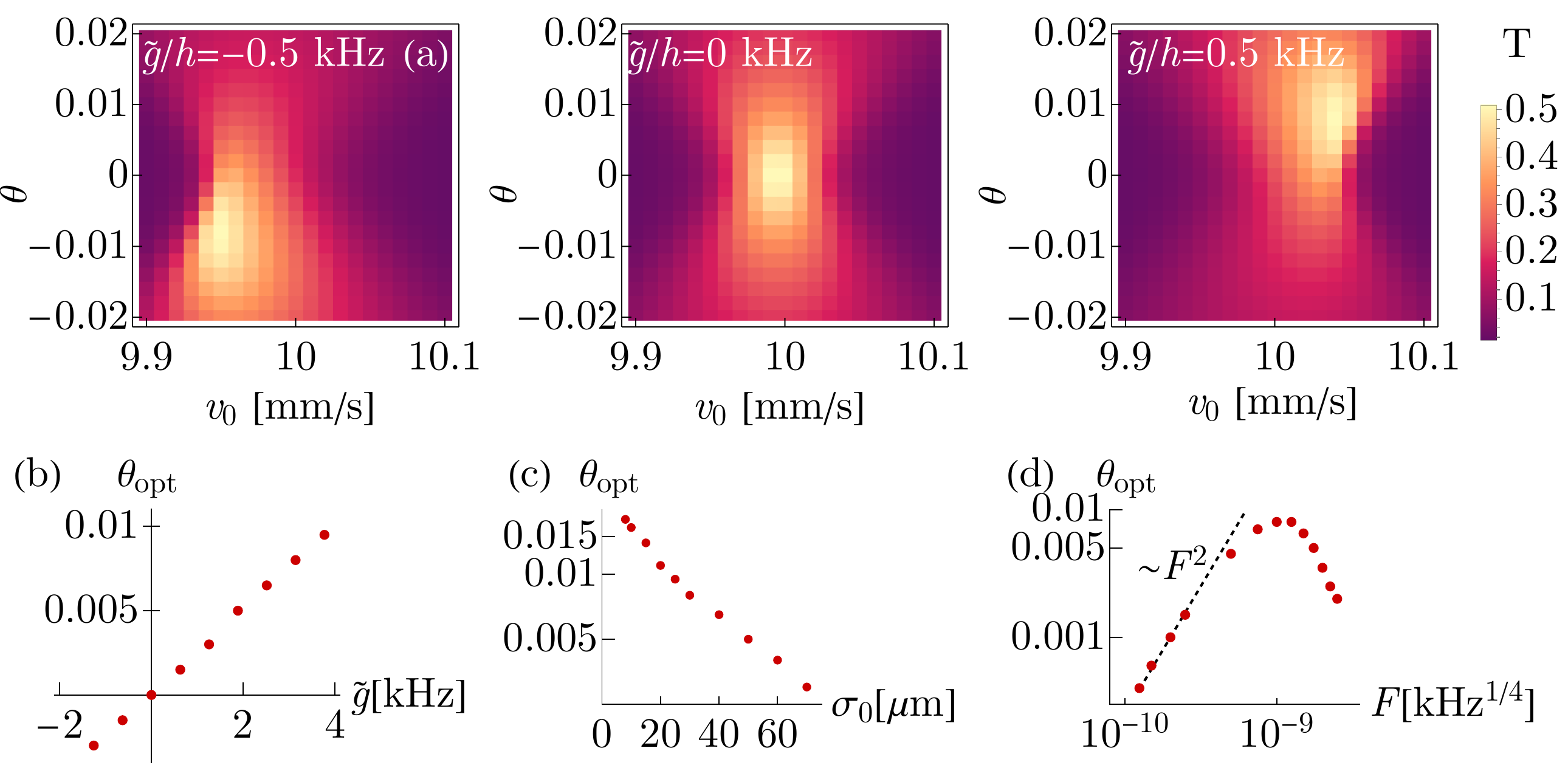}
\caption{(a) Transmission probability for different interaction strengths, chirp angles and initial velocities, obtained from our analytic approach via Eqs.~\eqref{eq_analytic},~\eqref{eq_transmissionAnalytic}. We set the cavity bound state energy to $E_{\mathrm{b}}/h = 11$ kHz, the cavity position to $x_0 = 250$ $\mu$m, the initial Gaussian state width to $\sigma_0 = 30$ $\mu$m, and the lensing harmonic oscillator length to $a_{\mathrm{ho}} = 2.3$ $\mu$m (corresponding to the same $\omega_{\mathrm{trap}}=22$ Hz used in earlier simulations). The scattering-bound coupling prefactor is set to $F = 10^{-9} \text{kHz}^{1/4}$ (the unit is due to our choice of energy normalization of scattering states, see Appendix~\ref{app_Analytics} for details). (b-d) Optimal chirp angle, obtained by maximizing transmission $T$ over all $v_0$ and $\theta$. We keep all parameters the same as panel (a), but vary (b) the renormalized interaction strength $\tilde{g}$, (c) the cloud size $\sigma_0$, and (d) the scattering-bound coupling strength $F$.}
\label{fig: analytics}
\end{figure}

\begin{equation}
\begin{aligned}
\label{eq_analytic}
i \hbar\frac{d}{dt} B(t) &= E_{\mathrm{b}}B(t) + \tilde{g} |B(t)|^2 B(t)-i \gamma(t)  B(t)\\
&+ P(t) e^{-i \mu(t) t/\hbar}.
\end{aligned}
\end{equation}
Here $E_{\mathrm{b}}$ is the energy of the bound state and $\tilde{g}$ is the interaction strength renormalized via $\tilde{g} = g \int dx |\phi(x)|^4$, with $\phi(x)$ the wavefunction of the bound state. The decay rate $\gamma(t)$ captures the coupling between bound and scattering states,
and is given by,
\begin{equation}
\label{eq_analyticGamma}
\gamma(t) = 2\pi V_{\mu(t)}^2.
\end{equation}
The energy-dependent coupling strength $V_{E}$ can be approximated as,
\begin{equation}
V_{E} = F E^{1/4},
\end{equation}
where $F$ is an overall constant encoding the barrier width, shape, and so on. The term $P(t) e^{-i \mu(t)t/\hbar}$ encodes the initial condition, which acts as a source driving the cavity state. The amplitude $P(t)$ should be thought of as the number of atoms arriving at time $t$, and $\mu(t)$ characterizes the energy of the atoms arriving at $t$. Crucially, that energy varies in time due to the (anti-) chirp. 

We take the analytically tractable case of a Gaussian-shaped initial wavepacket $\psi(x,0) = \frac{1}{\sqrt{\sigma_0}\pi^{1/4}}e^{-(x+x_0)^2/(2\sigma_0^2)}$ with spatial uncertainty $\sigma_0$ and distance $x_0$ from the cavity. To encode the chirp, this wavepacket is rotated in phase space by an angle $\theta$ in a harmonic trap of frequency $\omega_{\mathrm{trap}}$, hence oscillator length $a_{\mathrm{ho}} = \sqrt{\hbar / (m \omega_{\mathrm{trap}})}$, which we set to be much smaller than the cloud size $a_{\mathrm{ho}} \ll \sigma_0$. It is then given the usual momentum kick by multiplying the wavefunction by $e^{i k_0 x}$, giving it an initial energy $E_0 = \frac{\hbar^2 k_0^2}{2m}$. For this initial state, the source amplitude in the limit of small chirp angle $|\theta| \ll 1$ works out to be,
\begin{equation}
\begin{aligned}
P(t) &\approx P_0 \>\text{exp}\left[-\frac{E_0}{m \sigma_0^2}t^2\right],\\
P_0 &= 2F \mu_0^{1/4}\pi^{1/4}\sqrt{\frac{a_{\mathrm{ho}}}{\sigma_0}}\sqrt{E_0},
\end{aligned}
\end{equation}
where $\mu_0 = \hbar^2/(2m a_{\mathrm{ho}}^2) = \hbar \omega_{\mathrm{trap}}/2$, and we work in a frame where the packet's peak arrives at the cavity at time $t = 0$ (hence a packet initialized at distance $x_0$ from the cavity means an initial time of $t_0 = -x_0 / v_0$). The energy of the arriving atoms in the limit of $|\theta| \ll 1$ is,
\begin{equation}
\mu(t) \approx E_0+2E_0\mu_0\theta\frac{t}{\hbar},
\end{equation}
which is centered at the initial energy $E_0$ with a linear time-dependence set by the chirp angle. All approximations involved in this model and its derivation are detailed in Appendix~\ref{app_Analytics}.

Eq.~\eqref{eq_analytic} is a single scalar equation for the amplitude $B(t)$. While it is amenable to various analytic techniques, it can also be readily integrated numerically. Once the time-evolution of $B(t)$ is known, under this model we may obtain the transmission probability via:
\begin{equation}
\label{eq_transmissionAnalytic}
T = \frac{2\pi}{\hbar}\int_{-\infty}^{\infty}dt |B(t)|^2 V_{\mu(t)}^2.
\end{equation}

In Fig.~\ref{fig: analytics}(a) we plot the resulting transmission probability for different interaction strengths, initial velocities and chirp angles. The same physics found with our full GPE simulations is recovered: one sees a shifted, sharper resonant peak for a ``magic" anti-chirp for $g > 0$, and a corresponding chirp for $g < 0$.

Even without solving the equation of motion, one can also readily see the origin of the effect we consider. If we transform Eq.~\eqref{eq_analytic} into a frame rotating at the frequency of the arriving amplitude $B(t) \equiv B_s(t) e^{-i \mu(t) t/\hbar}$, for our choice of $\mu(t)$ we arrive at,
\begin{equation}
\begin{aligned}
i \hbar \frac{d}{dt}B_s(t) &= \bigg[E_{\mathrm{b}} - E_0- i \gamma(t)\\
&-4 E_0 \mu_0 \theta \frac{t}{\hbar} + \tilde{g} |B_s(t)|^2\bigg]B_s(t) + P(t).
\end{aligned}
\end{equation}
The terms inside the brackets on the last line contain a linear time-dependent frequency shift $4 E_0 \mu_0 \theta t/\hbar$ due to the chirp, as well as the mean-field shift $\tilde{g}|B_s(t)|^2$. If the latter grows linearly in time, there is a ``magic" choice of $\theta$ that cancels this growth when $4 E_0 \mu_0 \theta t/\hbar = \tilde{g}|B_s(t)|^2$. While we do not anticipate the mean-field shift to exhibit linear growth throughout the entire collision process (as the details depend on the shape of $P(t)$), the ``magic" chirp angle is the one that best cancels the linearized increase in mean-field shift at the time when the peak atomic amplitude is arriving.

From the transmission spectra in Fig.~\ref{fig: analytics}(a), we can also clearly see an optimal part of the spectrum where transmission is strongest. We can extract an optimal chirp angle $\theta_{\mathrm{opt}}$ by maximizing $T$ over all incident velocities $v_0$ and angles $\theta$ (for this scalar equation such maximization is numerically tractable). Figs.~\ref{fig: analytics}(b-d) report this optimal angle as a function of renormalized interaction strength $\tilde{g}$, initial cloud size $\sigma_0$, and scattering-bound coupling strength $F$ respectively. The $\tilde{g}$-scaling is linear, which is in line with the prediction of the chirp energy shift $\sim \theta$ cancelling the $\tilde{g}$-proportional mean-field shift. The optimal angle decreases roughly exponentially with cloud size $\sigma_0$ over the parameter range probed; the decrease is due to larger $\sigma_0$ implying smaller density and hence mean-field shift per unit time. The angle also grows as $\sim F^2$ for weak scattering-bound couplings, which is in line with the decay rate $\gamma$ from Eq.~\eqref{eq_analyticGamma}. For larger $F$, the optimal chirp angle instead decreases as the coupling grows so strong that transmission approaches 100\% regardless of incident velocity or chirp angle (i.e the collision becomes almost above-barrier for the resonance in question).

%%%%%%%%%%%%%%%%%%%%%%%%%%%%%%%%
\section{Conclusion}\label{Sec 5}
%%%%%%%%%%%%%%%%%%%%%%%%%%%%%%%%

We have shown that it is possible to resolve FP resonant features in BEC dynamics even in the presence of significant interactions. The core novel mechanism we describe is the use of position-momentum correlations of a wavepacket to dynamically cancel the density shift of a FP bound state due to atoms accumulating inside the cavity. This interaction-assisted tunneling enables applications such as near-discontinuous ``switching" of matter-wave transmission, akin to a low/high-pass filter for repulsive/attractive interatomic collisions. One can also employ the extreme sensitivity to initial velocity near the discontinuous part of a resonant peak for high-precision sensing, such as measurements of atomic acceleration due to gravity.

In addition to near-term experimental demonstration, future prospects of this work include extensions to more general forms of potential barriers such as optical lattices. Another alluring prospect is the dynamical tuning of cavity properties or interaction strength during collisional processes to trace out the hysteresis curve depicted in Fig.~\ref{fig: schematic}(c). If one can tune $g$ between separate collisions, one can even envision combining the low-pass and high-pass filter nature of the resonances to create very narrow momentum band-pass filters.

\section{ACKNOWLEDGMENTS}

This work has been supported by NSERC, the Fetzer Franklin Fund, and the QuEnSi quantum alliance under NSERC ALLRP 578468–22. We acknowledge helpful discussions with Joseph Thywissen and Nanako Shitara.

%\section*{Data Availability}

\bibliography{FPBibliography.bib}

\clearpage
\onecolumngrid
\appendix

%%%%%%%%%%%%%%%%%%%%%%%%%%%
\section{Experimental imperfections}
\label{app_Imperfections}
%%%%%%%%%%%%%%%%%%%%%%%%%%%

\begin{figure}
\centering
\includegraphics[width=0.6\textwidth]{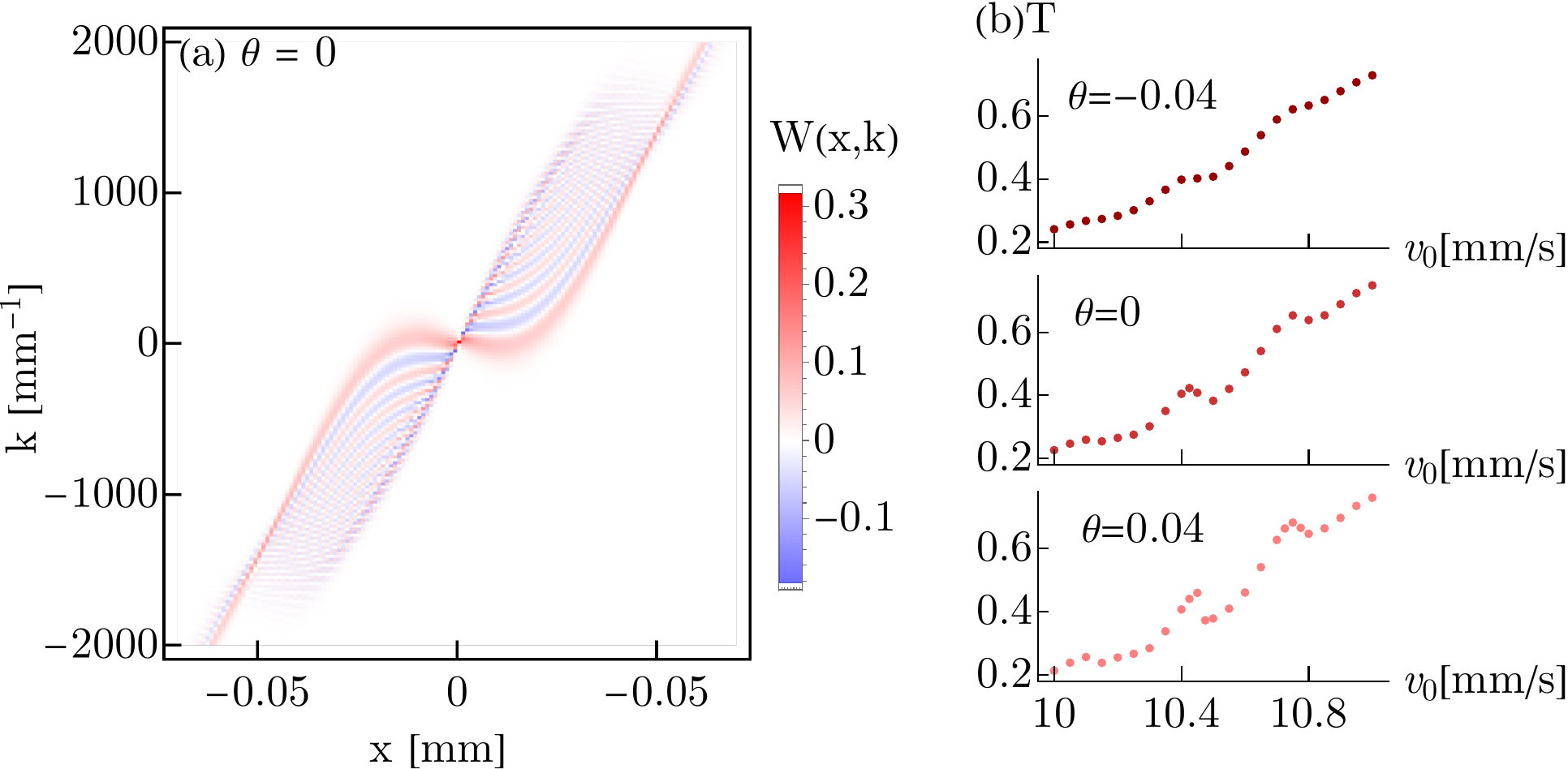}
    \caption{(a) Wigner function of the condensate after DKC for a anharmonic Gaussian trapping potential from Eq.~\eqref{eq_trapGaussian}. We use parameters $\sigma_{\mathrm{bw}}= 50$ $\mu$m, $\omega_{\mathrm{trap}}/h = 22$ Hz as before, and $t_{\mathrm{expand}} = 40$ ms of expansion time before applying the kick. The plotted state is for an ideal lensing time $t_{\mathrm{lens}} = t_0$ (hence chirping angle $\theta = 0$) that minimizes the initial variance $\Delta k$; for these parameters we find $t_0 = 1.85$ ms. Anharmonicity causes the outer fringes of the cloud to be under-rotated, compared to the ideal ellipse expected for a fully harmonic trap. (b) Transmission probabilities for a repulsive gas $g > 0$ and different chirp angles for imperfect states generated by the anharmonic trap. While overall resonance visibility is reduced, we still find a notable improvement for the right amount of anti-chirp.}
    \label{fig: expt}
\end{figure}

In this Appendix, we benchmark the effects of experimental imperfections on the visibility of FP resonances. The biggest source of error is anharmonicity in the harmonic potential $V_{\mathrm{trap}}(x)$ used for preparing the BEC and performing matter-wave lensing. A more realistic potential is created by a far-detuned laser beam which has a Gaussian spatial profile,
\begin{equation}
\label{eq_trapGaussian}
V_{\mathrm{trap,expt}}(x) = \frac{1}{2}m \omega_{\mathrm{trap}}^2 \frac{\sigma_{\mathrm{bw}}^2}{2}\left(1-e^{-\frac{2x^2}{\sigma_{\mathrm{bw}}^2}}\right),
\end{equation}
where $\sigma_{\mathrm{bw}}$ is the trap beam's waist ($1/e^2$ radius). One can verify this potential reduces to $\frac{1}{2}m \omega_{\mathrm{trap}}^2 x^2$ when Taylor expanded to second order in $x$. The anharmonicity will cause the portions of the BEC at larger $|x|$ to rotate more slowly during the matter-wave lensing. Instead of a perfect ellipse in phase space, one will get the profile shown in Fig.~\ref{fig: expt}(a).

Despite this under-rotation, we can still obtain an optimal lensing time $t_0$ such that the initial momentum uncertainty of the cloud is minimized, and under- or over-rotate by a chirp angle $\theta$. We can then evolve under the GPE with the FP cavity to extract transmission probabilities. Fig.~\ref{fig: expt}(b) shows the result; while the visibility of the resonances is reduced, there is still a clear advantage for the right amount of anti-chirp (here assuming repulsive $g > 0$ interactions).

We also note that one can get far better visibility by discarding the transmission probability coming from the outer ``lobes" of the cloud which are subject to significant anharmonicity, and only focusing on the central portion.

%%%%%%%%%%%%%%%%%%%%%%%%%%%
\section{Analytics}
\label{app_Analytics}
%%%%%%%%%%%%%%%%%%%%%%%%%%%

In this Appendix, we derive an effective scalar equation for the time-dependent population of the cavity during the collision. Several of these calculations follow and extend the treatment in Ref.~\cite{paul2007nonlinear}.

%%%%%%%%%%%%%%
\subsection{Equations of motion}
%%%%%%%%%%%%%%

Let's suppose the wavefunction of the condensate can be decomposed into a basis of one quasi-bound state inside the cavity (assuming other quasi-bound states are far enough in energy to not affect the dynamics), and unbound scattering continuum states left/right (L/R) of the cavity. We write the wavefunction as:
\begin{equation}
\psi(x,t) = \int_0^{\infty} dE \>A_{E}^{L}(t)\phi^{L}_{E}(x) +  B(t) \phi(x) + \int_0^{\infty}dE\> A_{E}^{R}(t)\phi_{E}^{R}(x).
\end{equation}
Here $\phi(x)$ is the quasi-bound eigenfunction with energy $E_{\mathrm{b}}$ and probability amplitude $B(t)$. Similarly, $\phi_{E}^{L/R}(x)$ are continuum eigenfunctions on the left/right side with energy $E$, and probability amplitudes $A_{E}^{L/R}(t)$. The continuum eigenfunctions are energy-normalized, meaning that $\int_{-\infty}^{\infty} dx [\phi_{E}^{L/R}(x)]^{*}\phi_{E'}^{L/R}(x) = \delta(E-E')$.

We insert this ansatz into the GPE. For the interaction term, we only keep diagonal contributions from the wavefunction inside the cavity, i.e. only a mean-field self-repulsion $\tilde{g} |B(t)|^2$, where the interaction strength is renormalized to be,
\begin{equation}
\tilde{g} = g \int dx |\phi(x)|^{4}.
\end{equation}
This approximation assumes that atomic density is much higher inside the cavity than outside, and that other bound states are sufficiently far in energy that we can't have interaction-mediated transitions.

We next assume that there is an energy-dependent coupling matrix element $V_E$ between a continuum state at energy $E$ on either side and the bound state. This leads to following equations for the wavefunction probability amplitudes:
\begin{equation}
\label{eq_EqOfMotion}
\begin{aligned}
i\hbar\frac{d}{dt} A_{E}^{L/R}(t) &= E A_{E}^{L/R}(t) + V_E B(t),\\
i\hbar\frac{d}{dt} B(t) &= E_{\mathrm{b}} B(t) + \tilde{g} |B(t)|^2 B(t)+\int_0^{\infty}dE \>V_E [A_{E}^{L}(t) + A_{E}^{R}(t)].
\end{aligned}
\end{equation}
As can be shown in Section~\ref{app_Analytics_Coupling}, a rough estimate of the coupling matrix element $V_E$ for our considerations is,
\begin{equation}
V_E \approx F E^{1/4},
\end{equation}
where $F$ is a constant that depends on details like the barrier height, barrier thickness, which resonance we are probing, and so on. Note that $V_E$ has units of $\sqrt{\text{Energy}}$, hence $F$ has units of $\text{Energy}^{1/4}$, due to our choice of energy-normalized continuum states.

The first line of Eq.~\eqref{eq_EqOfMotion} can be integrated in time:
\begin{equation}
\label{eq_ContinuumSol}
A_{E}^{L/R}(t) = A_{E}^{L/R}(t_0) e^{-i E (t-t_0)/\hbar} - \frac{i}{\hbar} V_E \int_{t_0}^{t}dt' B(t')e^{-i E (t-t')/\hbar}.
\end{equation}
We assume that there is only incoming amplitude from the left ($L$) side initially, and nothing on the right, hence $A_{E}^{R}(t_0)=0$. Inserting the equation for $A_{E}^{L/R}(t)$ into the second line of Eq.~\eqref{eq_EqOfMotion} gives:
\begin{equation}
\label{eq_NonlinearMotion}
i\hbar\frac{d}{dt} B(t) = E_{\mathrm{b}} B(t) + \tilde{g} |B(t)|^2 B(t)+ \int_{0}^{\infty} dE\>V_E A_{E}^{L}(t_0) e^{-iE(t-t_0)/\hbar} -\frac{2i}{\hbar}\int_{t_0}^{t}dt' B(t')\int_0^{\infty}dE \>V_E^2 e^{-i E(t-t')/\hbar}.
\end{equation}
This is a scalar equation for the cavity bound state amplitude $B(t)$ \textit{alone}, without any spatial degree of freedom, rendering the system much simpler to solve. In the following subsections, we will work to solve this equation by providing an explicit initial condition, simplifying the terms, and providing an explicit formula for the transmission probability.

%%%
\subsection{Initial condition}
%%%

We take the initial state at time $t_0$ to be a Gaussian with position uncertainty $\sigma_0$, center position $x_0$, finite velocity $v_0$ (hence momentum $k_0 =m v_0/\hbar$). A realistic cloud has a Thomas-Fermi profile, but we use a Gaussian for analytic simplicity. We also assume the wavefunction is \textit{rotated} about its center in phase space by a harmonic trap by a chirp angle $\theta \equiv \omega_{\mathrm{trap}}t_{\mathrm{lens}}$. The resulting wavefunction reads,
\begin{equation}
\label{eq_InitialWavefunctionAnalytic}
\begin{aligned}
\psi(x,t_0) &=\mathcal{N}_0 \text{exp}\left[-\frac{(x+x_0)^2}{2 a_{ho}^2}\Lambda+ik_0x\right],\\
\Lambda &= \frac{1- e^{-2i\theta}\text{tanh}\left[\text{ln}\left(\frac{\sigma_0}{a_{ho}}\right)\right]}{1+ e^{-2i \theta}\text{tanh}\left[\text{ln}\left(\frac{\sigma_0}{a_{ho}}\right)\right]},\\
\mathcal{N}_0 &= \frac{1}{\pi^{1/4}}\sqrt{\frac{\sigma_0}{\sigma_0^2 \cos(\theta)+ia_{ho}^2 \sin(\theta)}}.
\end{aligned}
\end{equation}
We want to use this initial state to obtain the initial amplitudes of the continuum states:
\begin{equation}
A_{E}^{L}(t_0) = \int_{-\infty}^{\infty} dx\> [\phi_{E}^{L}(x)]^{*} \psi(x,t_0).
\end{equation}
Far from the potential, the scattering wavefunctions should just look like travelling waves (the prefactor comes from the choice of energy normalization):
\begin{equation}
\label{eq_AsymptoticContinuum}
    \phi_{E}^{L}(x) = \frac{1}{2}\left(\frac{2m}{\pi^2 \hbar^2 E}\right)^{1/4}e^{i\frac{\sqrt{2m E}}{\hbar}x}.
\end{equation}
The initial amplitudes can now be evaluated:
\begin{equation}
\begin{aligned}
A_{E}^{L}(t_0) &= \mathcal{N}_0 \frac{1}{2}\left(\frac{2m}{\pi^2 \hbar^2 E}\right)^{1/4}\sqrt{\frac{2\pi}{\Lambda}}a_{ho} e^{i \frac{\sqrt{2m}}{\hbar}x_0(\sqrt{E}-\sqrt{E_0})-\frac{(\sqrt{E}-\sqrt{E_0})^2}{2\mu_0 \Lambda}},\\
\mu_{0} &= \frac{\hbar^2}{2m a_{\mathrm{ho}}^2},\>\>\>\>
\end{aligned}
\end{equation}
where $\mu_{0}$ is akin to an energy uncertainty, and $E_0 = \frac{\hbar^2 k_0^2}{2m}$ is the initial mean kinetic energy of the wavepacket.

%%%%%%%%%%%%%%%%%%%
\subsection{Source amplitude}
%%%%%%%%%%%%%%%%%%%

Armed with these amplitudes, we can compute the third term on the right hand side of Eq.~\eqref{eq_NonlinearMotion}. Recalling that we have $V_E \approx F E^{1/4}$ for some constant $F$, we get:
\begin{equation}
\begin{aligned}
\int_0^{\infty} dE\>V_E A_{E}^{L}(t_0)e^{-iE(t-t_0)/\hbar}&=F\mathcal{N}_0 \frac{1}{2}\left(\frac{2m}{\pi^2 \hbar^2}\right)^{1/4}\sqrt{2\pi \Lambda}\>a_{ho} \mu_0 e^{-i \frac{\sqrt{2m}}{\hbar}x_0\sqrt{E_0}-\frac{E_0}{2\mu_0 \Lambda}}\\
&\quad\quad\quad\quad\times\bigg[\frac{2}{Q(t)}+\sqrt{2\pi}e^{\frac{R^2}{2Q(t)}}\frac{R}{Q(t)^{3/2}}\left(1+\text{erf}\left[\frac{R}{\sqrt{2Q(t)}}\right]\right)\bigg],\\
R&=\sqrt{\frac{E_0}{\mu_0 \Lambda}}+ i \frac{\sqrt{2m}}{\hbar}x_0\sqrt{\mu_0 \Lambda},\\
Q(t) &= 1+2i\mu_0 \Lambda\frac{t-t_0}{\hbar}.
\end{aligned}
\end{equation}
For the system parameters we consider, it is generally a good approximation that $|R(t)| \gg |\sqrt{Q(t)}|$. This is because $R(t)$ contains the factor of $\sqrt{E_0/(\mu_0 \Lambda)}$ with $E_0$ the initial energy and $\mu_0 \Lambda$ characterizing uncertainty in that energy; hence $|R(t)| \gg 1$ when the initial energy is much larger than its uncertainty. By comparison, $|Q(t)|$ is of order 1 unless the elapsed time $(t-t_0)/\hbar$ times the uncertainty $\mu_0 \Lambda$ is much larger than 1 (i.e. the initial uncertainty has had sufficient time to act), by which time the collision has generally already finished. We can thus Taylor expand the expression in $R(t)/\sqrt{Q(t)}$ and obtain a simpler form:
\begin{equation}
\int_0^{\infty} dE\>V_E A_{E}^{L}(t_0)e^{-iE(t-t_0)/\hbar}\approx F\mathcal{N}_0 \left(\frac{2m}{\pi^2 \hbar^2}\right)^{1/4}2\pi \sqrt{\Lambda} a_{ho} \mu_0\frac{R}{Q(t)^{3/2}} e^{-i \frac{\sqrt{2m}}{\hbar}x_0\sqrt{E_0}-\frac{E_0}{2\mu_0 \Lambda}+\frac{R^2}{2Q(t)}}.
\end{equation}
Re-inserting the definitions of $R$ and $Q(t)$ and simplifying, we find:
\begin{equation}
\begin{aligned}
\int_0^{\infty} dE\>V_E A_{E}^{L}(t_0)e^{-iE(t-t_0)/\hbar}&\approx F \mathcal{N}_0 \mu_{0}^{3/4} 2\sqrt{\pi a_{\mathrm{ho}}\Lambda} \frac{\sqrt{\frac{E_0}{\mu_0 \Lambda}}+i \frac{\sqrt{2m}}{\hbar}x_0 \sqrt{\mu_0 \Lambda}}{\left(1+2i \mu_0 \Lambda\frac{t-t_0}{\hbar}\right)^{3/2}}\\
&\times\text{exp}\left[-\frac{\mu_0 \Lambda}{2}\frac{\left(2\sqrt{E_0}\frac{t-t_0}{\hbar}-\frac{\sqrt{2m}}{\hbar}x_0\right)^2}{1+2i \mu_0 \Lambda\frac{t-t_0}{\hbar}}\right] e^{-i E_0 \frac{t-t_0}{\hbar}}.
\end{aligned}
\end{equation}

While the above expression is still quite ugly, we can make it more simple with more assumptions about our initial condition. Recall that $\Lambda$ in Eq.~\eqref{eq_InitialWavefunctionAnalytic} involves the factor $\tanh\left[\text{ln}\left(\frac{\sigma_0}{a_{\mathrm{ho}}}\right)\right]$ (a ratio of the harmonic trap width used for matter-wave lensing, and the initial cloud size). If we assume a broad initial cloud and a tight initial trap with $\sigma_0 \gg a_{\mathrm{ho}}$, we can Taylor expand the expression in $\sigma_0 / a_{\mathrm{ho}}$. Under this approximation, we can write the expression as,
\begin{equation}
\int_0^{\infty} dE\>V_E A_{E}^{L}(t_0)e^{-iE(t-t_0)/\hbar}\approx P(t) e^{-i \Delta(t)} + \mathcal{O}\left(\frac{a_{\mathrm{ho}}}{\sigma_0}\right),\\
\end{equation}
where the amplitude and phase are given by:
\begin{equation}
\begin{aligned}
P(t) &=
F \mathcal{N}_0 \mu_0^{1/4} 2\sqrt{\pi a_{\mathrm{ho}}}\frac{\sqrt{E_0}-\frac{\sqrt{2m}}{\hbar}x_0 \mu_0\tan(\theta)}{\left[1 -2 \mu_0 \tan(\theta)\frac{t-t_0}{\hbar}\right]^{3/2}}\text{exp}\left[-\frac{\hbar^2\sec^2(\theta)}{4m\sigma_0^2}\frac{\left(2\sqrt{E_0}\frac{t-t_0}{\hbar}-\frac{\sqrt{2m}}{\hbar}x_0\right)^2}{1-2 \mu_0 \tan(\theta)\frac{t-t_0}{\hbar}}\right], \\
\Delta(t) &=E_0\frac{t-t_0}{\hbar} - \frac{\left(2\sqrt{E_0}\frac{t-t_0}{\hbar} -\frac{\sqrt{2m}}{\hbar}x_0\right)^2}{2\left(2\frac{t-t_0}{\hbar}-\frac{\cot(\theta)}{\mu_0}\right)},
\end{aligned}
\end{equation}
The amplitude $P(t)$ is real-valued up to a constant phase, and has a maximum at,
\begin{equation}
t_{\mathrm{max}}-t_0 = \frac{\sqrt{m}x_0}{\sqrt{2E_0}} = \frac{m x_0}{\hbar k_0} = \frac{x_0}{v_0},
\end{equation}
which is just the time needed for the wavepacket to cross the distance $x_0$ with initial velocity $v_0$. In the limit of small chirp angle $|\theta| \ll 1$, the amplitude simplifies to,
\begin{equation}
P(t) = 2 \pi^{1/4} F \mu_0^{1/4}\sqrt{\frac{a_{\mathrm{ho}}}{\sigma_{0}}} \sqrt{E_0} \exp \left[-\frac{E_0}{m \sigma_0^2}(t-t_{\mathrm{max}})^2\right] + \mathcal{O}(\theta).
\end{equation}

Up to an overall constant, we can re-write the phase as,
\begin{equation}
\begin{aligned}
\Delta(t) &= \mu(t) \frac{t-t_{\mathrm{max}}}{\hbar},\\
\mu(t) &= E_0 - \frac{2 E_0}{2\frac{t-t_0}{\hbar}-\frac{1}{\mu_0}\cot(\theta)}\frac{t-t_{\mathrm{max}}}{\hbar}.
\end{aligned}
\end{equation}
Assuming the chirp angle to be small $|\theta| \ll 1$, the function $\mu(t)$ can be simplified:
\begin{equation}
\mu(t) = E_0 + 2 E_0 \mu_0 \theta \frac{t-t_{\mathrm{max}}}{\hbar} + \mathcal{O}(\theta^2).
\end{equation}
Notice this $\mu(t)$ contains the ``bare" mean energy of the wavepacket $E_0$, plus a time-linear-dependent term that captures the effect of the chirp through the lensing angle $\theta$. In the main text, we shift time by $t_{\mathrm{max}}$, so that the wavepacket arrives at time $t = 0$ for simplicity.

%%%%%%%%%%%%%%%%%
\subsection{Kernel}
%%%%%%%%%%%%%%%%%

Having determined the time-dependent contributions of the source term, we turn to the last term of Eq.~\eqref{eq_NonlinearMotion}. We know that at a given time $t$, our source is driving the system at center frequency $e^{-i \mu(t) t/\hbar}$. The frequency $\mu(t)$ varies with time, but slowly compared to other timescales in the system. We can write that last term of Eq.~\eqref{eq_NonlinearMotion} as:
\begin{equation}
-\frac{2i}{\hbar} \int_{t_0}^{t} dt'B(t') \int_{0}^{\infty} dE\> V_E^2 e^{-i E (t-t')/\hbar} = -\frac{2i}{\hbar} \int_{t_0}^{t} dt' B(t')e^{-i \mu(t')(t-t')/\hbar}\int_{0}^{\infty} dE\> V_E^2 e^{-i(E-\mu(t'))(t-t')/\hbar}.
\end{equation}
In the above, we effectively transform the integrand with respect to $t'$ into a frame rotating at the mean frequency of the source term at time $t$. We now make the approximation that $B(t') e^{-i \mu(t')(t-t')}$ varies slowly with $t'$, compared to the energy integral, because we've already divided out by the fast-oscillating source frequency. Since the energy integral only gets appreciable contributions from times $t'$ close to the current time $t$ (acting as a memory kernel), we can approximate $B(t')e^{-i \mu(t')(t-t')} \approx B(t)$, thereby assuming the time integrand is roughly constant on timescales relevant to the energy integral. We further assume the initial time $t_0$ to be sufficiently far back in time that we can send it to $t_0 \to - \infty$. This leaves us with:
\begin{equation}
-\frac{2i}{\hbar} \int_{t_0}^{t} dt'B(t') \int_{0}^{\infty} dE\> V_E^2 e^{-i E (t-t')/\hbar} \approx -\frac{2i}{\hbar} B(t)\int_{-\infty}^{t} dt' \int_{0}^{\infty} dE\> V_E^2 e^{-i(E-\mu(t))(t-t')/\hbar}.
\end{equation}
We exchange order of integration, and apply the Dirac delta function (ignoring the Cauchy principal value contribution, which should only give a small correction to our final result):
\begin{equation}
\begin{aligned}
\int_{-\infty}^{t} dt' \int_{0}^{\infty} dE\> V_E^2 e^{-i(E-\mu(t))(t-t')/\hbar} &= \int_{0}^{\infty} dE\> V_E^2 \int_{-\infty}^{t} dt' e^{-i(E-\mu(t))(t-t')/\hbar}\\
&=\int_{0}^{\infty} dE\>V_E^2  \int_{0}^{\infty}d\tau \>e^{-i (E-\mu(t))\tau/\hbar}\\
&=\hbar\int_0^{\infty}dE \>V_E^2 \pi \delta(E-\mu(t))\\
&=\hbar\pi V_{\mu(t)}^2.
\end{aligned}
\end{equation}
Our equation for $B(t)$ now, finally, becomes:
\begin{equation}
i\hbar\frac{d}{dt}B(t) = E_{\mathrm{b}}B(t) + \tilde{g} |B(t)|^2 B(t) + P(t) e^{-i \mu(t) t/\hbar}-2\pi i V_{\mu(t)}^2B(t),
\end{equation}
The time-dependence of $\mu(t)$ is crucial in this equation: it captures the fact that at different times, atoms impact the cavity with different energies due to the chirping angle.

%%%%%%%%%%%%%%%%%
\subsection{Transmission probability}
%%%%%%%%%%%%%%%%%

Since Ref.~\cite{paul2007nonlinear} considered the quasistationary case, they could directly solve for time-independent $B(t)$. We are considering a transient problem, so we still have to solve the equation of motion for $B(t)$ numerically. But this is now a scalar equation, which can be solved very easily and quickly.

Once we (numerically) solve for $B(t)$, we can also obtain the outbound amplitude from Eq.~\eqref{eq_ContinuumSol}:
\begin{equation}
\begin{aligned}
A_{E}^{R}(t) &= -\frac{i}{\hbar}V_E \int_{-\infty}^{t}dt' B(t') e^{-i E(t - t')/\hbar}.
\end{aligned}
\end{equation}
The transmission probability is then given by the total integral of the outgoing amplitude:
\begin{equation}
\begin{aligned}
T &= \int_0^{\infty} dE |A_E^R{(t \to \infty)}|^2\\
&= \frac{1}{\hbar^2}\int_0^{\infty} dE \> V_E^2 \int_{-\infty}^{\infty}dt' \int_{-\infty}^{\infty}dt'' B^{*}(t')B(t'')e^{-i E(t'-t'')/\hbar}.
\end{aligned}
\end{equation}
We can simplify this integral by playing the same tricks as before:
\begin{equation}
T = \frac{1}{\hbar^2}\int_0^{\infty} dE \>V_E^2 \int_{-\infty}^{\infty}dt' B^{*}(t')\int_{-\infty}^{\infty}dt''B(t'')e^{-i \mu(t'')(t'-t'')/\hbar}\>e^{-i (E-\mu(t''))(t'-t'')/\hbar}. 
\end{equation}
We approximate $B(t'')e^{-i \mu(t'')(t'-t'')/\hbar} \approx B(t')$:
\begin{equation}
\label{eq_FinalTransmission}
\begin{aligned}
T &\approx \frac{1}{\hbar^2}\int_0^{\infty} dE \>V_E^2 \int_{-\infty}^{\infty}dt' |B(t')|^2 \int_{-\infty}^{\infty}dt''\>e^{-i (E-\mu(t'))(t'-t'')/\hbar}\\
&=\frac{1}{\hbar^2}\int_0^{\infty} dE\>V_E^2 \int_{-\infty}^{\infty}dt' |B(t')|^2 \int_{-\infty}^{\infty}d\tau e^{-i(E-\mu(t'))\tau/\hbar}\\
&=\frac{1}{\hbar^2}\int_0^{\infty} dE\>V_E^2 \int_{-\infty}^{\infty}dt' |B(t')|^2\>2\pi \hbar \delta(E-\mu(t'))\\
&=\frac{2\pi}{\hbar}\int_{-\infty}^{\infty}dt' |B(t')|^2 V_{\mu(t')}^2.
\end{aligned}
\end{equation}

%%%%%%%%%%%%%%%%%%%
\subsection{Coupling between continuum and bound states}\label{app_Analytics_Coupling}
%%%%%%%%%%%%%%%%%%%

We want to evaluate the coupling $V_E$ between scattering states and quasi-bound cavity states.

We consider the simplified situation where the cavity barriers are square potentials of thickness $l$ each:
\begin{equation}
V(x) =\begin{cases}
0 & |x| > \frac{d}{2} + l\\
V_0 &  \frac{d}{2} < |x| < \frac{d}{2}+l\\
0 & |x| < \frac{d}{2}
\end{cases}
\end{equation}
This potential hosts a finite number $N_b$ of quasi-bound states. If we assume the barriers to be sufficiently thick and ignore the empty space outside them, this is simply a finite square well problem. The allowed energies are determined implicitly from the equations:
\begin{equation}
\label{eq_boundImplicit}
\kappa = k \tan\left(\frac{k d}{2}\right),\>\>\>\>\>\> \kappa = - k \cot \left(\frac{k d}{2}\right),
\end{equation}
where $k$ and $\kappa$ are,
\begin{equation}
k = \sqrt{\frac{2mE}{\hbar^2}},\>\>\>\>\>\kappa = \sqrt{\frac{2m(V_0-E)}{\hbar^2}}.
\end{equation}
A real-valued solution to the first/second equation in Eq.~\eqref{eq_boundImplicit} for the energy $E$ is a valid bound state with even/odd parity about $x=0$ respectively. We index these bound states by $n= 0, 1, \dots N_b-1$ with energies $E_n$ and corresponding momenta $k_n$, $\kappa_n$, ordered by increasing energy hence alternating parity. So $n=0,2,4,\dots$ are even-parity and $n=1,3,5\dots$ are odd-parity. The corresponding wavefunctions can be written as,
\begin{equation}
\psi_n(x) = \begin{cases}
2\mathcal{N}_n k_n e^{\kappa_n \left(x+\frac{d}{2}\right)} & x < -\frac{d}{2}\\
\mathcal{N}_n \left[(k_n-i\kappa_n)e^{i k_n \left(x + \frac{d}{2}\right)}+(k_n + i \kappa_n)e^{-i k_n \left(x + \frac{d}{2}\right)}\right] & |x| \leq \frac{d}{2} \\
\mathcal{N}_n \left[(k_n -i\kappa_n)e^{i k_n d} +(k_n + i \kappa_n)e^{-i k_n d}\right]e^{-\kappa_n \left(x - \frac{d}{2}\right)} & x > \frac{d}{2}
\end{cases}
\end{equation}
where $\mathcal{N}_n$ is given by,
\begin{equation}
\mathcal{N}_n = \left[\frac{\kappa_n}{2\left(k_n^2+\kappa_n^2\right)\left(2+\kappa_n d\right)}\right]^{1/2} = \left[\frac{\hbar^2 \sqrt{2m(V_0-E_n)}}{4m V_0 \left(2\hbar + d \sqrt{2m(V_0 - E_n)}\right)}\right]^{1/2}.
\end{equation}

We next compute the wavefunctions of scattering continuum states outside the cavity. Let's do the ones on the left side of the cavity. If we assume the left cavity barrier is just a step potential, thereby assuming the scattering state gets attenuated before it reaches the inside of the cavity, the scattering-state wavefunction is easy enough to write:
\begin{equation}
\phi^{L}_{k}(x) = \begin{cases}
\mathcal{N}_{k}\left[(k - i \kappa)e^{i k \left(-x-\frac{d}{2}-l\right)} + (-k+i\kappa) e^{-i k \left(x-\frac{d}{2}-l\right)}\right] & x \leq -\frac{d}{2} - l\\
2\mathcal{N}_{k} k e^{\kappa \left(-x-\frac{d}{2}-l\right)} & x > -\frac{d}{2} - l
\end{cases}
\end{equation}
where we have distinguished the scattering momentum $k$ and corresponding imaginary momentum $\kappa = \sqrt{\frac{\hbar^2 V_0}{2m}-k^2}$ from the $k_n$, $\kappa_n$ of the cavity bound states.

To connect with the derivations in the previous sections, we want our continuum eigenfunctions to be energy-normalized. As shown in Eq.~\eqref{eq_AsymptoticContinuum} earlier, far from any potential these should look like travelling-waves $\phi_{E}^{L}(x) \approx\frac{1}{2}\left(\frac{2m}{\pi^2 \hbar^2 E}\right)^{1/4} e^{i \frac{\sqrt{2m E}}{\hbar} x}$, or alternatively trigonometric functions $\phi_{E}^{L}(x) \approx\left(\frac{2m}{\pi^2 \hbar^2 E}\right)^{1/4} \cos(\frac{\sqrt{2m E}}{\hbar}x)$. We re-write our eigenfunctions as a single cosine,
\begin{equation}
\phi^{L}_{k}(x) = \begin{cases}
2\mathcal{N}_{k}\sqrt{k^2+\kappa^2}\cos\left[k\left(-x-\frac{d}{2}-l\right)-\text{arctan}\left(\frac{\kappa}{k}\right)\right]& x \leq -\frac{d}{2} - l\\
2\mathcal{N}_{k} k e^{\kappa \left(-x-\frac{d}{2}-l\right)} & x > -\frac{d}{2} - l
\end{cases}
\end{equation}
By comparing the prefactor of the first line to $\left(\frac{2m}{\pi^2 \hbar^2 E}\right)^{1/4}$, we can thus read off:
\begin{equation}
\mathcal{N}_k =\frac{1}{\sqrt{k^2+\kappa^2}}\left(\frac{m}{8\pi^2 \hbar^2 E}\right)^{1/4}.
\end{equation}

Now, we can obtain a coupling matrix element. The simplest way to do this is with a Bardeen surface integral,
\begin{equation}
V_E = -\frac{\hbar^2}{2m}\left[(\phi_{k}^{L})^{*}(x)\frac{\partial \psi_n(x)}{\partial x} - \psi_n(x) \frac{\partial (\phi_k^L)^{*}(x)}{\partial x}\right]_{x= -\frac{d}{2}-\frac{l}{2}}.
\end{equation}
namely, the probability current from the scattering to the quasi-bound wavefunction evaluated at the center of the left barrier. Evaluating this matrix element gives us:
\begin{equation}
\begin{aligned}
V_E &= -\frac{\hbar^2}{2m}2\frac{k}{\sqrt{k^2+\kappa^2}}\left(\frac{m}{8 \pi^2 \hbar^2 E}\right)^{1/4} 2\mathcal{N}_{n}k_{n} \left(e^{\kappa(-x-\frac{d}{2}-l)}\frac{\partial e^{\kappa_n(x+\frac{d}{2})}}{\partial x}-e^{\kappa_n(x+\frac{d}{2})}\frac{\partial e^{\kappa (-x-\frac{d}{2}-l)}}{\partial x}\right)\bigg|_{x=-\frac{d}{2}-\frac{l}{2}}\\
&=-2\left(\frac{2 m}{\pi^2 \hbar^2 V_0^2}\right)^{1/4}\mathcal{N}_n \sqrt{E_n} (\sqrt{V_0-E}+\sqrt{V_0-E_n}) e^{-l\frac{ \sqrt{m}}{\sqrt{2}\hbar}\left(\sqrt{V_0-E}+\sqrt{V_0 - E_n}\right)}E^{1/4}.
\end{aligned}
\end{equation}
Assuming the scattering energy is close to the bound state energy $E \approx E_n$, we can simplify further to,
\begin{equation}
\begin{aligned}
V_E &\approx F E^{1/4},\\
F &= -4\left(\frac{2 m}{\pi^2 \hbar^2 V_0^2}\right)^{1/4}\mathcal{N}_n \sqrt{E_n} \sqrt{V_0 - E_n}e^{-l\frac{ \sqrt{2m(V_0-E_n)}}{\hbar}}.
\end{aligned}
\end{equation}
The factor $F$ encodes all other information about the resonance. It's worth noting that we do not expect this formula for $F$ to be quantitatively correct for our scattering results, because i) we consider smooth barriers rather than steps, and ii) we consider scattering for states with energy very close to the barrier height (or even above it). Nonetheless, we can get good estimates by just setting an $F$ by hand, and seeing what transmission spectra emerge.

\end{document}